\documentclass{article}
\usepackage{spconf,amsmath,graphicx,hyperref}

\usepackage{cite}
\usepackage{spconf, hyperref,amssymb,amsfonts,amsthm}
\usepackage{algorithmic}
\usepackage{graphicx}
\usepackage{textcomp}
\usepackage{xcolor}
\usepackage{NotationMacros}
\usepackage{layouts}
\usepackage{subcaption}
\usepackage{url}
\usepackage{booktabs}
\usepackage{tikz}
\usetikzlibrary{arrows.meta, positioning, shapes.geometric, fit, calc}

\title{AnyRIR: Robust Non-intrusive Room Impulse Response Estimation in the Wild}

\name{Kyung Yun Lee$^{1}$\thanks{KYL and SJS were supported through the joint German Academic Exchange Service (DAAD) and Research Council of Finland (RCF) Project (57763119) "Immersive Augmented Acoustics (IAA)".} \quad Nils Meyer-Kahlen$^{1}$ \quad Karolina Prawda$^{2}$ \quad Vesa Välimäki$^{1}$ \quad Sebastian J. Schlecht$^{3}$}
  
\address{$^{1}$ Aalto University \\
      $^{2}$ University of York \\ 
      $^{3}$ Friedrich-Alexander-Universitat Erlangen-Nürnberg (FAU)}
      
\begin{document}
\ninept
\maketitle

\begin{abstract}
We address the problem of estimating room impulse responses (RIRs) in noisy, uncontrolled environments where non-stationary sounds such as speech or footsteps corrupt conventional deconvolution. We propose AnyRIR, a non-intrusive method that uses music as the excitation signal instead of a dedicated test signal, and formulate RIR estimation as an $\ell_1$-norm regression in the time–frequency domain. Solved efficiently with  Iterative Reweighted Least Squares (IRLS) and Least-Squares Minimal Residual (LSMR) methods, this approach exploits the sparsity of non-stationary noise to suppress its influence. Experiments on simulated and measured data show that AnyRIR outperforms $\ell_2$-based and frequency-domain deconvolution, under in-the-wild noisy scenarios and codec mismatch, enabling robust RIR estimation for AR/VR and related applications.
\end{abstract}

\begin{keywords}
Acoustic measurements, deconvolution, music.
\end{keywords}

\section{Introduction}
\label{sec:intro}

A room impulse response (RIR) characterizes the acoustic properties of an environment, capturing how sound propagates and reflects within a space. Given a specific RIR, it is possible to render arbitrary sound sources as if they were played within that particular environment \cite{Shoichi2025}. This capability is especially valuable for applications in augmented and virtual reality (AR/VR) and smart speakers, where realistic simulation of various acoustic spaces, from domestic spaces to large public venues, is essential \cite{gupta2022armr, herre2023mpeg}. 

\begin{figure}[bt!]
    \centering
    \includegraphics[trim=0cm 0cm 0cm 0cm, clip,width=0.9\linewidth]{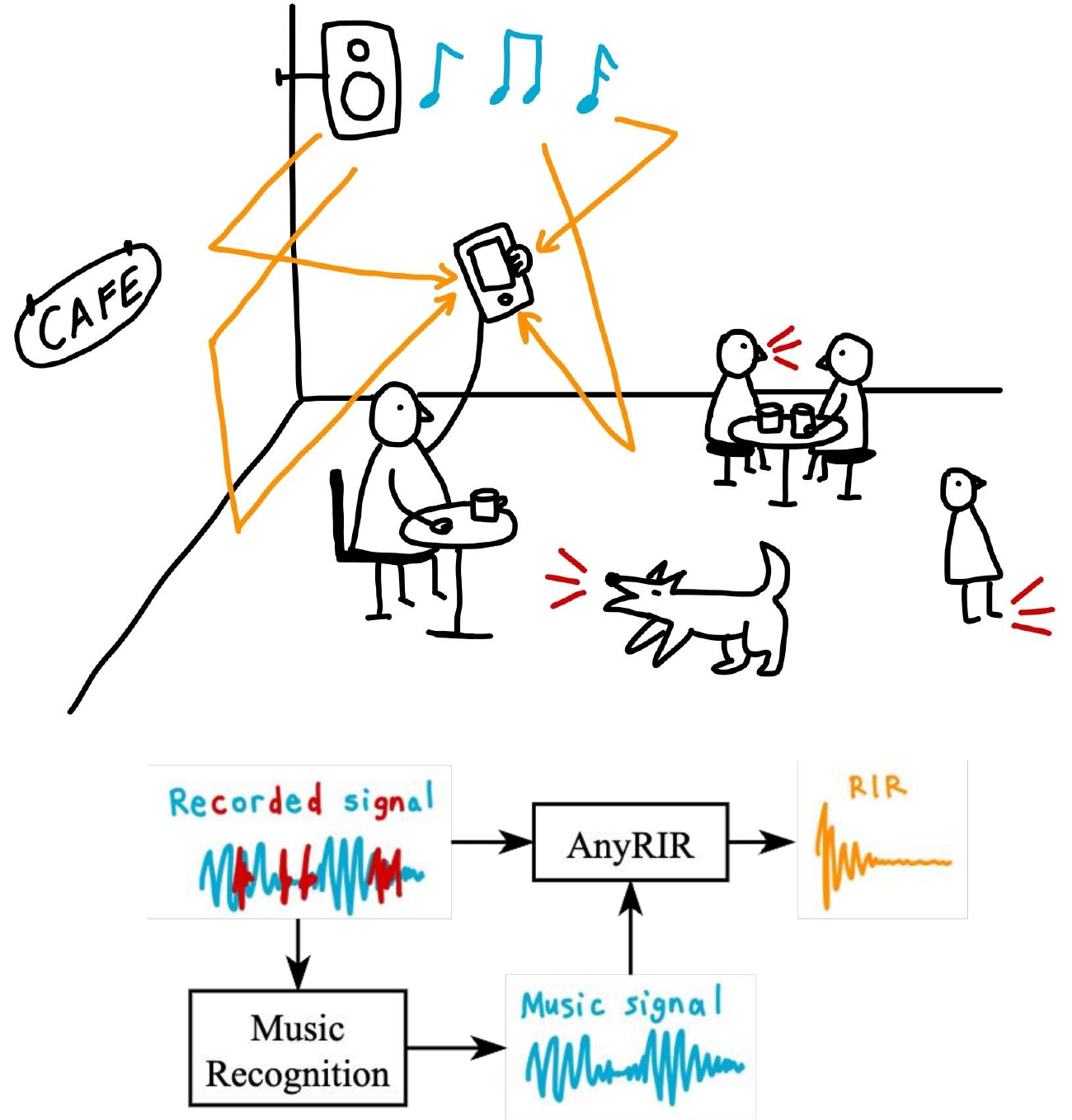}
    \caption{(Top) A noisy environment, such as cafe, where music is playing and is reverberating, and (bottom) usage of AnyRIR to estimate the RIR with the help of existing music recognition algorithm.}
    \label{fig:AnyRIRprinciple}
\end{figure}

However, estimating RIRs in uncontrolled real-world environments remains a challenge \cite{Shoichi2025}. Conventional methods, such as exponential sine sweeps, require controlled and often intrusive setups that are impractical in public or dynamic spaces. Moreover, non-stationary noise sources (e.g., footsteps or speech) corrupt these measurements \cite{ciric2011, Prawda:2022.Sweep, prawda2024non}. Inspired by the widespread presence of background music in public venues and the availability of clean references through music identification algorithms such as Shazam \cite{shazam2025, wang_industrial-strength_2003}, we propose a robust and non-intrusive method for estimating RIRs in the wild using music as the excitation signal. Fig.~\ref{fig:AnyRIRprinciple} illustrates how our method can estimate an RIR from music recorded in a cafe.

While stationary noise with a Gaussian distribution can often be handled adequately by $\ell_2$-based deconvolution methods, non-stationary noise violates the underlying assumptions and severely degrades estimation performance \cite{stan2002comparison}. MOSAIC \cite{prawda2024non} introduces a way to remove non-stationary noise from exponential sine sweep measurements using median filtering, but it requires at least three repeated measurements with exactly the same excitation signal, which makes it somewhat inflexible. As an alternative, we adopt a regression-based approach using the $\ell_1$-norm loss, which is known for its robustness to outliers \cite{Boyd:2004jj}. This makes it particularly suitable for handling non-stationary noise, since it suppresses the influence of transient disturbances in the measurement. Although $\ell_1$-norm regularization has previously been used to promote sparsity in RIR estimation \cite{crocco2015room}, inducing sparsity in the filter is not our objective. Instead, the $\ell_1$-norm is used solely as the data fidelity term to improve robustness to non-stationary noise, without imposing sparsity or other constraints on the RIR.

\begin{figure*}[ht]
    \centering
    \includegraphics[trim=0cm 0cm 0cm 0cm, clip,width=1.0\textwidth]{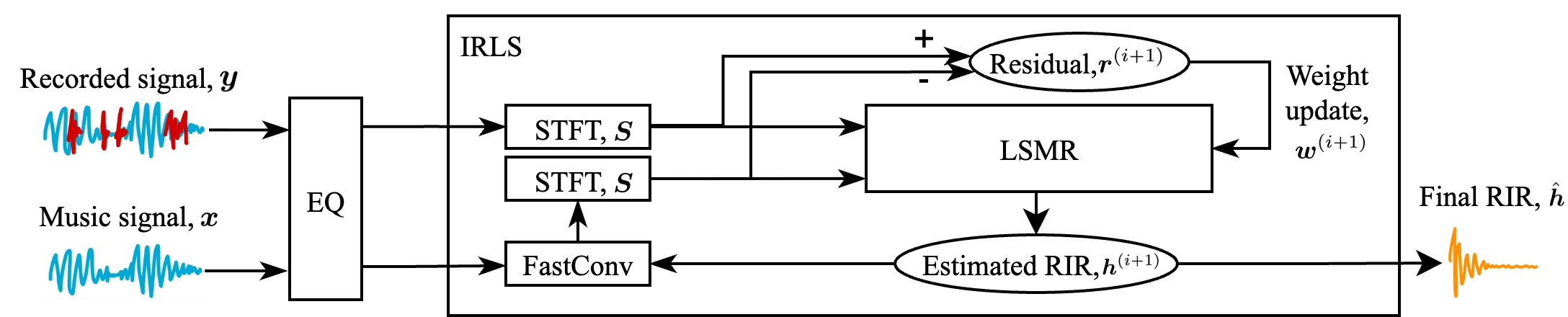}
    \caption{The proposed AnyRIR method estimates the RIR by minimizing the time-frequency residual between the recorded music and the clean music convolved with the previous RIR estimate. The main tools used are the IRLS and LSMR methods.}
    \label{fig:AnyRIRalgorithm}
\end{figure*}

Robust deconvolution using the $\ell_1$-norm has been explored in other domains. In underwater acoustics, it has been applied to channel estimation under impulsive noise using time-domain ADMM \cite{Cang.2022}. In image processing, $\ell_1$-norm minimization has been widely used to suppress impulsive noise during deconvolution \cite{Bar_2007_deconvolution, cho2011handling}. In contrast, our approach operates in the time–frequency domain and does not impose sparsity on the filter, focusing instead on robustness to non-stationary noise in real-world acoustic measurements.

\section{Problem Formulation}


Given an RIR $\vec{h}$ and an excitation signal $\vec{x}$, the measured signal $\vec{y}$ at a microphone can be modeled as $\vec{y} = \vec{x} \conv \vec{h} + \vec{n},$ where $\conv$ denotes the convolution operation and $\vec{n}$ represents additive background noise. Let $\vec{x}$ have length $L$, and $\vec{h}$ have length $N$, then $\vec{y}$ has length $M = L + N -1$. 
We aim to solve the inverse problem of estimating the unknown RIR $\vec{h}$ from a known pair of excitation and measurement signals, $\vec{x}$ and $\vec{y}$. When the noise $\vec{n}$ is assumed to be Gaussian, modeling stationary background noise, the deconvolution problem can be approached with the $\ell_2$-norm loss. However, real-world noise often contains non-stationary impulsive components, which is better modeled by heavy-tailed, non-Gaussian distributions. Such impulsive components act as outliers in the residual and degrade the performance of $\ell_2$-based estimators. To enhance robustness against outliers, we adopt the $\ell_1$-norm loss \cite{Boyd:2004jj} and write the convolution as a multiplication operator by forming the Toeplitz matrix $\mat{X}$ of $\vec{x}(t)$:  
\begin{align}
\begin{aligned}
    \hat{\vec{h}} &= \argmin_{\vec{h}} \norm{\vec{y} - \vec{x} \conv \vec{h}}_1 = \argmin_{\vec{h}} \norm{\vec{y} - \mat{X} \vec{h} }_1.
\end{aligned}
\label{eq:this}
\end{align}

\noindent Furthermore, many non-stationary disturbances are sparse in the time-frequency domain, e.g., speech \cite{Rickard.2002, Plumbley.2010}. Thus, we perform a time-frequency residual computation: 
\begin{align}
\begin{aligned}
    \hat{\vec{h}} = \argmin_{\vec{h}} \norm{ \mat{S} \vec{y} - \mat{S} \mat{X} \vec{h}}_1,
\label{eq:l1_norm}	
\end{aligned}
\end{align}
where $\mat{S}$ is the time-frequency transform matrix, e.g., the stacked matrix of DFT matrices which forms the short-time Fourier transform (STFT) matrix.

\section{Proposed Estimation Method}

The proposed method, AnyRIR, solves the deconvolution problem using the $\ell_1$-norm minimization (\ref{eq:l1_norm}) in the time-frequency domain, where $\mat{X}$ and $\mat{S}$ are large matrices. This leads to a time and memory limitation when using standard optimization libraries, such as CVXPY \cite{diamond2016cvxpy}. There are various efficient alternatives, such as ADMM \cite{Cang.2022}. In particular, we choose the Iterative Reweighted Least Squares (IRLS) method along with a matrix-free iterative method called the Least-Squares Minimal Residual (LSMR) to handle the least-squares subproblem  \cite{fong2011lsmr}, as shown in Fig.~\ref{fig:AnyRIRalgorithm}, while avoiding to form $\vec{X}$ and $\vec{S}$ explicitly.
The IRLS algorithm is formulated as a sequence of weighted least squares problems: 
\begin{align}
\begin{aligned}
    \vec{h}^{(i+1)} &= \argmin_{\vec{h}} \norm{ \vec{w}^{(i)} \hadamard \left (\mat{S} (\vec{y} - \mat{X} \vec{h}^{(i)})\right )}_2, \\
    \vec{r}^{(i+1)} &= \mat{S} (\vec{y} - \mat{X} \vec{h}^{(i+1)}), \\
    \vec{w}^{(i+1)} &= \frac{1}{\max( \lvert(\vec{r}^{(i+1)}\rvert, \delta^{(i)})},
\label{eq:IRLS}	
\end{aligned}
\end{align}
\noindent with the initial weight matrix $\vec{w}^{(0)} = \vec{1}$, initial estimate $\vec{h}^{(0)} = \vec{0}$. The symbol $\hadamard$ denotes the Hadamard product, $\lvert . \rvert$ taking the absolute value and max$()$, refers to elementwise selection of the larger of its two arguments.  The weight update scheme corresponds to a Huber loss function \cite{Boyd:2004jj}, where $\delta$ defines the threshold between quadratic and linear behavior. In our setting, where both stationary and non-stationary noise are present, $\delta$ can be interpreted as the estimated standard deviation of the background noise. Residuals with magnitude below this threshold are assumed to arise from stationary noise and are given full weight, while larger residuals, which are assumed to be outliers due to transient or non-stationary interference, are downweighted to $1/|\vec{r}|$, thus improving robustness in the presence of outliers.

For small-scale problems where $\mat{X}$ is well-conditioned, the least squares subproblem can be solved analytically using the normal equations. However, in our setting, where the excitation signal is very long, forming the normal equation becomes memory-intensive. In addition, the convolutional structure of $\mat{X}$ often leads to ill-conditioned systems \cite{chan1994circulant}, resulting in numerical instability. Therefore, we employ an iterative solver, such as the Conjugate Gradient method \cite{Shewchuk.1994} and LSQR \cite{paige1982lsqr}. In particular, we use LSMR, which is a matrix-free method for sparse and or fast linear operator, offering improved robustness when handling ill-conditioned matrices \cite{fong2011lsmr}. 

LSMR requires both a forward and an adjoint operator, as it does not compute the normal equation explicitly. The forward operator computes $ \vec{y} = \mat{S} \mat{X} \vec{h}$, a matrix-form of convolution operator followed by a time-frequency transform. The adjoint operator reverses this sequence, $\vec{h} = \mat{X}^{H}\mat{S}^{H} \vec{y}$, where $\mat{S}^{H}$ is the adjoint of the time-frequency transform and $\mat{X}^{H}$ is the adjoint of convolution, corresponding to the time-reversed excitation signal.

Both convolution and its adjoint are implemented using FFT-based operations, avoiding explicit construction of the convolution matrix $\mat{X}$. The FFT-based convolution and the adjoint convolution are computed as 
\begin{align}
\begin{aligned}
    \mat{X} \vec{h} &= \textrm{trunc}_N (\textrm{iFFT}_M (\textrm{FFT}_M(\vec{x}) \hadamard \textrm{FFT}_M(\vec{h})))\\ 
    \mat{X}\herm \vec{y} &= \textrm{trunc}_N (\textrm{iFFT}_M (\textrm{FFT}_M(\textrm{flip}(\vec{x})) \hadamard \textrm{FFT}_M(\vec{y}))),
\label{eq:this}	
\end{aligned}
\end{align}

\noindent where $M$ is the FFT length and $N$ is the output length. Time-frequency transform $\mat{S}$ and its Hermitian $\mat{S}^{H}$ are implemented with the SciPy STFT and iSTFT functions, which are linear, properly paired, and normalized such that the iSTFT acts as the adjoint of the STFT. To achieve this, we apply a boxcar window with no overlap, use zero-padding of $2 N_\mathrm{DFT}$, and apply a normalization factor of $\sqrt{N_\mathrm{DFT}/2}$, since the scipy STFT uses an unnormalized FFT. We fix the DFT length to $N_\mathrm{DFT} = 256$. 

\begin{figure}[t!]
    \centering
\includegraphics[trim=0cm 0cm 0cm 0cm, clip, width=0.9\linewidth]{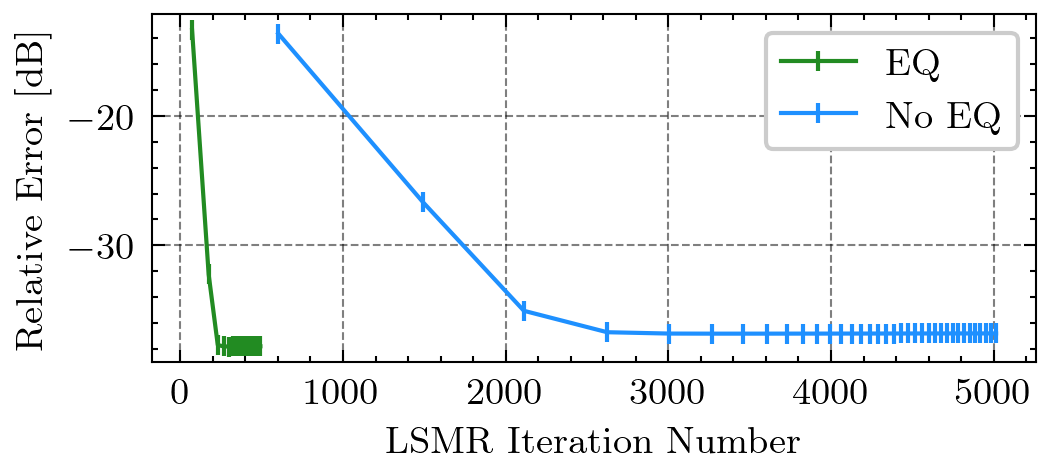}
\caption{Number of LSMR iterations required for convergence with and without EQ preconditioning.}
    \label{fig:convergence}
\end{figure}

The conditioning of the system matrix $\vec{X}$ is critical for the convergence of iterative solvers in least squares problems. In our case, the system matrix $\vec{X}$ is a Toeplitz matrix. For such a matrix to be well-conditioned, the power spectral density (PSD) should be approximately flat, i.e., the excitation signal should be spectrally white. However, music signals typically do not have a flat spectrum. While preconditioning techniques for such Toeplitz matrices are well studied \cite{chan1994circulant}, flattening the PSD of the excitation signal achieves a similar effect. We therefore apply preprocessing to equalize (EQ) both the excitation and measurement signals, as shown in Fig.~\ref{fig:AnyRIRalgorithm}. Specifically, we downsample the audio to 32\,kHz, inject high-frequency noise, and apply a shared high-order inverse linear prediction filter (order 200). This step acts as a preconditioner, improving numerical stability and convergence of IRLS-based deconvolution. Fig.~\ref{fig:convergence} presents the improvement in convergence, showing that the EQ preconditioning reduces the number of necessary iterations by a factor of 10. The code for an implementation of the full method is available online\footnote{\url{https://github.com/kyungyunlee/robust-deconv}}.

\section{Evaluation}

We evaluate the proposed method in both simulated and real-world scenarios that reflect typical use cases of AnyRIR, such as in cafés. Our approach is compared against two baselines: an $\ell_{2}$-norm minimizing method in the time-frequency domain, which assumes only Gaussian noise, and a frequency-domain deconvolution method \cite{farina2000simultaneous}. Furthermore, we assess the robustness of AnyRIR under codec mismatch conditions, where the music played in the environment differs in encoding from the reference identified by the music recognition algorithm, a situation that may frequently occur in practice.

\subsection{Dataset}
To simulate music played in a typical public environment, we convolve music signals with RIRs and add both stationary and non-stationary noise. Specifically, we generated four different songs using Suno AI\cite{suno2025}, covering styles characteristic of café environments: pop, melodic acoustic instrumentation, coffee-shop ambiance, male and female vocals, light percussion, and full-band arrangements. For non-stationary noise, we use the AID dataset \cite{goetz_AID_2022}, which contains anechoic recordings of real-world interferers such as footsteps or coughs. RIRs are from the MIT Acoustical Reverberation Scene Statistics Survey \cite{MIT_survey}, which contains 271 RIR representing everyday acoustic environments.


\subsection{Presence of Non-stationary Noise}

  


  

\begin{figure*}[ht]
    \centering
    \includegraphics[trim=0cm 0cm 0cm 0cm, clip, width=0.9\textwidth]{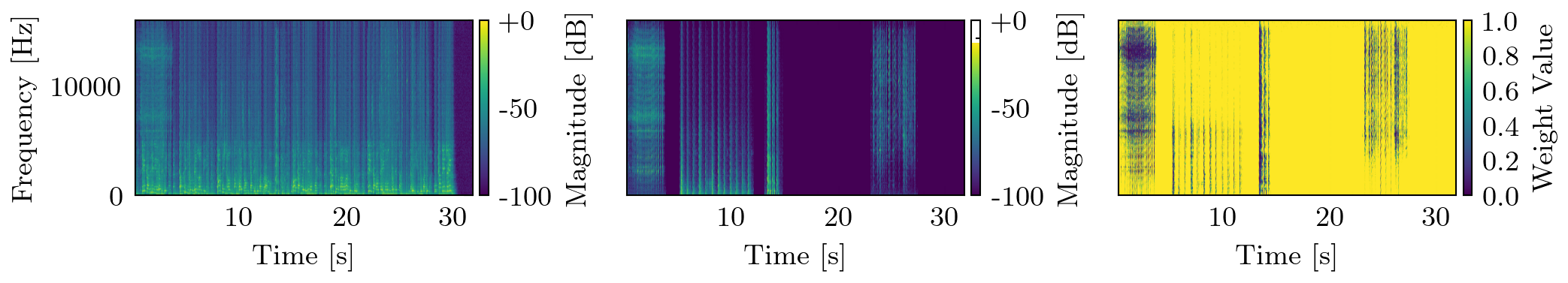}
    \caption{(Left) Music signal convolved with an RIR with added stationary and non-stationary noise: $\vec{y}$. (Center) Original non-stationary noise. (Right) Weights $\vec{w}$ computed by AnyRIR to suppress the bins containing the non-stationary noise.}
    \label{fig:spectrograms}
\end{figure*}

\begin{figure}[ht!]
    \centering
    \includegraphics[trim=0cm 0cm 0cm 0cm, clip,width=0.9\columnwidth]{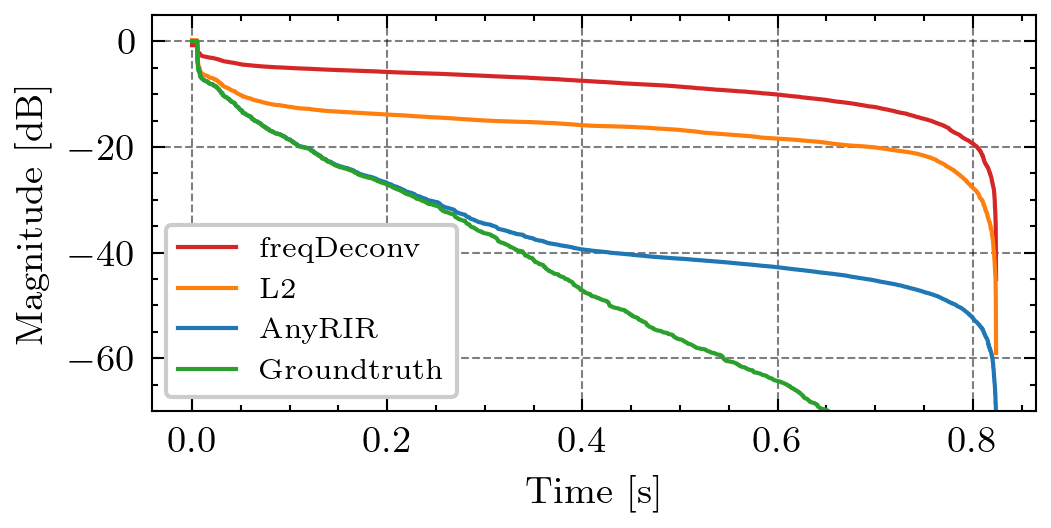}
    \caption{EDCs from different deconvolution methods compared to the ground truth, showing the superiority of AnyRIR.}
    \label{fig:main_result}
\end{figure}

We evaluate the robustness of AnyRIR on the task of deconvolution in the presence of non-stationary noise. For this experiment, we generated 50 synthetic examples. Each example consists of a 30-s music excerpt convolved with an RIR, with non-stationary noise events (coughing, cutlery, footsteps, glass jar) inserted at random times without overlap. The total duration of non-stationary noise per sample is between 20–50\% of the music signal. Further, stationary noise is added to achieve a fairly high SNR of 50\,dB, since stationary background-noise suppression is outside the scope of this work.

The method is designed to treat time–frequency bins containing non-stationary noise as outliers and down-weight them during RIR estimation. Fig.~\ref{fig:spectrograms} shows an example signal $\vec{x}$ alongside the detected non-stationary noise and the corresponding weights $\vec{w}$ assigned by AnyRIR. As seen by comparing the middle and right panels, the method successfully identified noise-dominated bins and suppressed them with weights close to zero, while preserving the clean signal and stationary background noise with weights near one. This behavior reflects the design of our weighting scheme, which attenuates large deviations while leaving small fluctuations unaffected.
The effect on RIR estimation can be seen in Fig.~\ref{fig:main_result}, which compares the corresponding energy decay curves (EDCs). Relative to the ground-truth RIR, AnyRIR’s estimate deviates only slightly due to background noise in the late tail. In contrast, the $\ell_{2}$-norm and frequency-domain deconvolution methods produce substantially poorer estimates.

As an upper bound, we consider performance on signals without non-stationary noise. Stationary noise remains challenging to suppress, but its effect can be averaged out by extending measurement signals, e.g., to 2–3 min. Table~\ref{tab:h_error} shows that AnyRIR performs equivalently to the standard $\ell_{2}$ method in the absence of non-stationary noise, while providing robustness when such noise is present. Results are reported as mean and standard deviation across the 50 examples.

\begin{table}[t!]
\small
\centering
\caption{Comparison of RIR estimation error ($\vec{h}$ error in dB, mean $\pm$ std) under different noise conditions. AnyRIR maintains low error in both cases, while baseline methods degrade substantially in the presence of non-stationary noise.}
\begin{tabular}{l|c|c|c}
\toprule
 & \textbf{AnyRIR} & \textbf{L2} & \textbf{Freq. Deconv.} \\ 
\midrule
\textbf{Stat. noise only} & $-42.0 \pm 4.8$ & $-41.7 \pm 4.8$ & $-7.6 \pm 4.9$ \\
\textbf{Stat. + non-stat.} & $-36.0 \pm 5.0$ & $-10.6 \pm 6.8$ & $2.8 \pm 4.5$ \\
\bottomrule
\end{tabular}

\label{tab:h_error}
\end{table}

\subsection{Music Codec}

In real-world scenarios, the music played through a loudspeaker may use a different codec than the reference track retrieved by a music recognition algorithm, e.g., one in WAV, the other in MP3. We call this a \emph{codec mismatch} scenario, which effectively introduces additional background noise.

To simulate this condition, we degraded signals using FFmpeg by encoding and decoding with different codecs. We considered two cases: in the \textit{codec mismatch} condition, the measurement signal was compressed with MP3 at 173 kbps while the excitation signal was compressed at 64 kbps; in the \textit{same codec} condition, both signals were taken from the same MP3 file. This setup isolates the effect of codec differences during system identification.

\begin{figure}[t!]
    \centering
    \includegraphics[trim=0cm 0cm 0cm 0cm, clip,width=0.9\columnwidth]{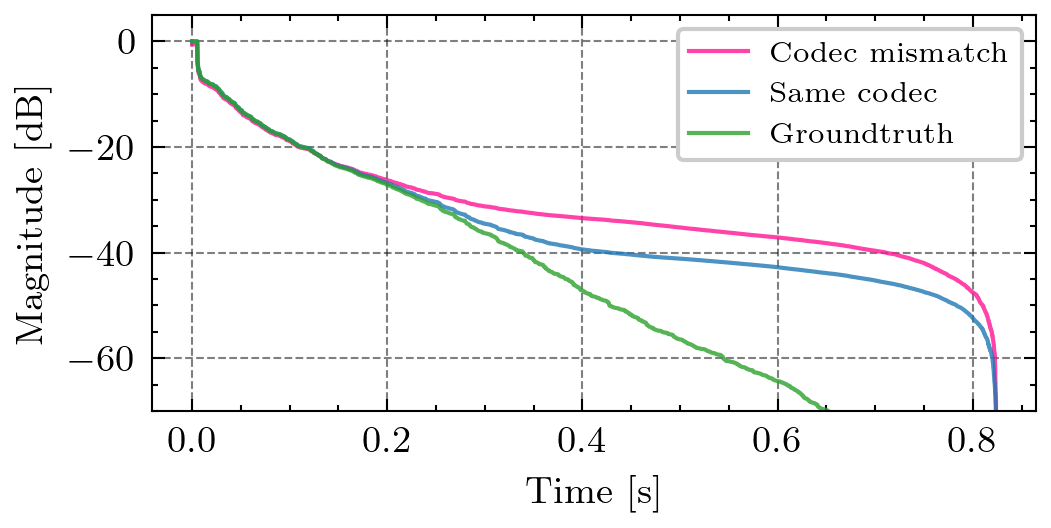}
    \caption{Effect of codec mismatch. When the measurement and excitation signals use different codecs, the estimated RIR exhibits increased noise in the late tail, resulting in $\sim$15 dB higher error compared to the same-codec condition.}
    \label{fig:codec_mismatch}
\end{figure}

Figure~\ref{fig:codec_mismatch} shows that codec mismatches primarily affect the late tail of the estimated RIR, where increased noise is observed. Quantitatively, the mismatch condition leads to approximately 15 dB higher error compared to the same-codec condition ($-22$ dB vs. $-37$ dB $\vec{h}$ error). This degradation is consistent with the interpretation of codec mismatch as an effective background noise source.

\subsection{Real-world Evaluation}

To evaluate AnyRIR under real-world conditions, we conducted measurements in the kitchen space of the Aalto Acoustics Lab. The ground-truth RIR was obtained using the exponential sine sweep method. For evaluation, we played a 1-min AI-generated music excerpt through a Genelec 8020A loudspeaker and recorded the response with an AKG C414 XLS microphone.

\begin{figure}[t!]
\centering
\includegraphics[trim=0cm 0cm 0cm 0cm, clip,width=0.9\columnwidth]{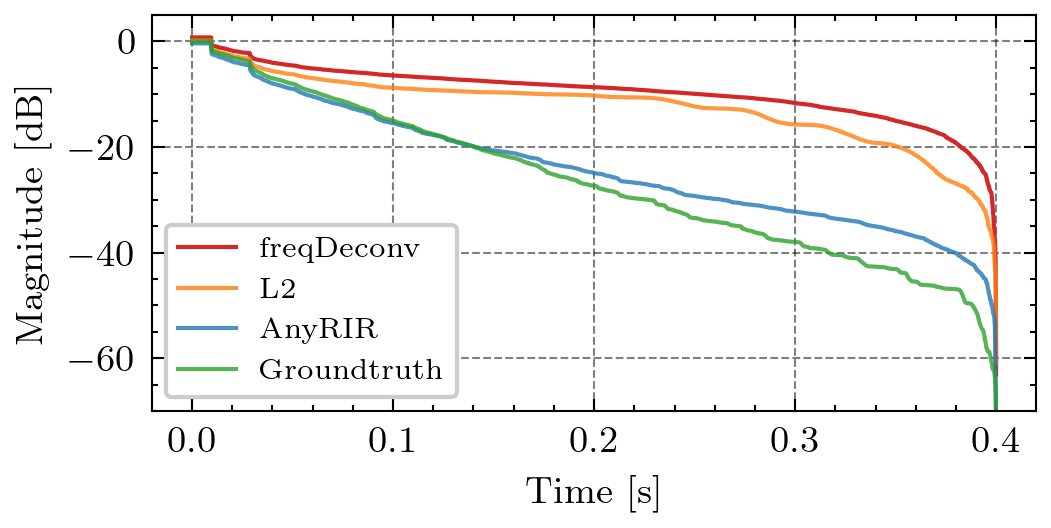}
\caption{EDC estimated from a real-world recording. }
\label{fig:real-world}
\end{figure}

During playback, lab members casually generated everyday noises such as door slams, cutlery sounds, and chair movements. We collected three recordings with varying noise densities, two of which included speech. Fig.~\ref{fig:real-world} shows the EDCs of one of the recordings, where AnyRIR provides an estimate close to the ground truth despite the presence of interfering noise. The related audio examples, video, and additional results are available online\footnote{\url{https://kyungyunlee.github.io/anyRIR-demo}}.

\section{Conclusions}
This paper introduced AnyRIR, a robust method for estimating RIRs from music as a non-intrusive alternative to exponential sine sweeps. To handle long excitation signals and non-stationary noise, we cast the problem as an $\ell_{1}$-norm optimization, solved efficiently with IRLS and LSMR in the time–frequency domain and further accelerated using FFT-based convolution. Experiments on 50 simulated recordings and three real-world measurements showed that AnyRIR yields accurate estimates and outperforms $\ell_{2}$-norm and frequency-domain deconvolution in the presence of non-stationary noise and codec mismatch.

\newpage
\bibliographystyle{IEEEbib}
\bibliography{IEEEexample, My_Library, Nils_Library_additions}

@inproceedings{cho2011handling,
  title={Handling outliers in non-blind image deconvolution},
  author={Cho, Sunghyun and Wang, Jue and Lee, Seungyong},
  booktitle={Proc. Int. Conf. Computer Vision},
  pages={495--502},
  year={2011},
}

@inproceedings{goetz_AID_2022,
  title={{AID}: Open-source anechoic interferer dataset},
  author={G{\"o}tz, Philipp and Tuna, Cagdas and Walther, Andreas and Habets, Emanu{\"e}l AP},
  booktitle={Proc. Int. Workshop on Acoustic Signal Enhancement (IWAENC)},
  pages={1--5},
  year={2022},
}

@article{chan1994circulant,
  title={Circulant preconditioned {Toeplitz} least squares iterations},
  author={Chan, Raymond H and Nagy, James G and Plemmons, Robert J},
  journal={SIAM J. Matrix Analysis and Applications},
  volume={15},
  number={1},
  pages={80--97},
  year={1994},
  publisher={SIAM}
}

@article{paige1982lsqr,
  title={{LSQR}: An algorithm for sparse linear equations and sparse least squares},
  author={Paige, Christopher C and Saunders, Michael A},
  journal={ACM Trans. Mathematical Software (TOMS)},
  volume={8},
  number={1},
  pages={43--71},
  year={1982},
  publisher={ACM New York, NY, USA}
}

@article{fong2011lsmr,
  title={{LSMR}: An iterative algorithm for sparse least-squares problems},
  author={Fong, David Chin-Lung and Saunders, Michael},
  journal={SIAM J. Scientific Computing},
  volume={33},
  number={5},
  pages={2950--2971},
  year={2011},
  publisher={SIAM}
}

@inproceedings{farina2000simultaneous,
  title={Simultaneous measurement of impulse response and distortion with a swept-sine technique},
  author={Farina, Angelo},
  booktitle={Proc. AES 108th Convention},
  year={2000},
  month={Feb.},
  note={paper no.~5093}
}

@article{prawda2024non,
  title={Non-stationary noise removal from repeated sweep measurements},
  author={Prawda, Karolina and Schlecht, Sebastian J and V{\"a}lim{\"a}ki, Vesa},
  journal={JASA Express Lett.},
  volume={4},
  number={8},
  year={2024},
  eprint ={081601},
  month={Aug.},
}

@INPROCEEDINGS{ciric2011,
    author={D. {Ćirić} and A. {Pantić} and D. {Radulović}},
    booktitle={Proc. 10th Int. Conf. Telecommunication in Modern Satellite Cable and Broadcasting Services (TELSIKS)}, 
    title={Transient noise effects in measurement of room impulse response by swept sine technique}, 
    year={2011},
    pages={269-272},
    month={Oct.},}

@article{stan2002comparison,
  title={Comparison of different impulse response measurement techniques},
  author={Stan, Guy-Bart and Embrechts, Jean-Jacques and Archambeau, Dominique},
  journal={J. Audio Eng. Soc.},
  volume={50},
  number={4},
  pages={249--262},
  year={2002},
}

@article{Bar_2007_deconvolution,
author = {Bar, Leah and Sochen, Nir and Kiryati, Nahum},
title = {Convergence of an Iterative Method for Variational Deconvolution and Impulsive Noise Removal},
journal = {Multiscale Modeling \& Simulation},
volume = {6},
number = {3},
pages = {983-994},
year = {2007},
doi = {10.1137/060671607}}

@article{MIT_survey,
author = {James Traer  and Josh H. McDermott },
title = {Statistics of natural reverberation enable perceptual separation of sound and space},
journal = {Proc. Natl. Acad. Sci. U.S.A.},
volume = {113},
number = {48},
pages = {E7856--E7865},
year = {2016},
doi = {10.1073/pnas.1612524113}}

@INPROCEEDINGS{Shoichi2025,
  author={Koyama, Shoichi and De Sena, Enzo and Samarasinghe, Prasanga and Thomas, Mark R. P. and Antonacci, Fabio},
  booktitle={Proc. IEEE Int. Conf. Acoust. Speech  Signal Process. (ICASSP)}, 
  title={Past, Present, and Future of Spatial Audio and Room Acoustics}, 
  year={2025},
  month={Apr.},
  volume={},
  number={},
  pages={1-5}}

@article{gupta2022armr,
	author = {Gupta, Rishabh and He, Jianjun and Ranjan, Rishabh and Gan, Woon-Seng and Klein, Florian and Schneiderwind, Christian and Neidhardt, Annika and Brandenburg, Karlheinz and V{\"a}lim{\"a}ki, Vesa},
	journal = {IEEE Signal Process. Mag.},
	month = {May},
	number = {3},
	pages = {63-89},
	title = {Augmented/Mixed Reality Audio for Hearables: {S}ensing, control, and rendering},
	volume = {39},
	year = {2022}}

@article{herre2023mpeg,
	author = {Herre, J{\"u}rgen and Disch, Sascha},
	journal = {J. Audio Eng. Soc.},
	month = {May},
	number = {5},
	pages = {229--240},
	title = {{MPEG-I} Immersive Audio---{R}eference Model For The Virtual/Augmented Reality Audio Standard},
	volume = {71},
	year = {2023}}

@article{Prawda:2022.Sweep, 
year = {2022}, 
title = {{Robust selection of clean swept-sine measurements in non-stationary noise}}, 
author = {Prawda, Karolina and Schlecht, Sebastian J and Välimäki, Vesa}, 
journal = {J. Acoust. Soc. Am.}, 
issn = {0001-4966}, 
doi = {10.1121/10.0009915}, 
pmid = {35364928}, 
pages = {2117--2126}, 
number = {3}, 
volume = {151}, 
}

@inproceedings{Cang.2022, 
year = {2022}, 
title = {Robust Deconvolution of Underwater Acoustic Channels Corrupted by Impulsive Noise}, 
author = {Cang, Siyuan and Sheng, Xueli and Jakobsson, Andreas and Yang, Huayong}, 
booktitle = {Proc. 5th Int. Conf. Info. Comm. Signal Process. (ICICSP)}, 
pages = {571--576}
}

@inproceedings{Rickard.2002, 
year = {2002}, 
title = {On the Approximate {W}-Disjoint Orthogonality of Speech}, 
author = {Rickard, Scott and Yilmaz, {\"O}.}, 
booktitle = {Proc. IEEE Int. Conf. Acoust. Speech Signal Process. (ICASSP)}, 
pages = {529-532}, 
volume = {1},
}

@article{Plumbley.2010, 
year = {2010}, 
title = {Sparse Representations in Audio and Music: From Coding to Source Separation}, 
author = {Plumbley, Mark D. and Blumensath, Thomas and Daudet, Laurent and Gribonval, RÃ©mi and Davies, Mike E.}, 
journal = {Proc. IEEE}, 
issn = {0018-9219}, 
doi = {10.1109/jproc.2009.2030345}, 
pages = {995--1005}, 
number = {6}, 
volume = {98}
}

@book{Shewchuk.1994, 
year = {1994}, 
title = {{An Introduction to the Conjugate Gradient Method Without the Agonizing Pain}}, 
author = {Shewchuk, Jonathan Richard}
}

@article{Boyd:2004jj, 
year = {2004}, 
title = {{Convex Optimization}}, 
author = {Boyd, Stephen P and Vandenberghe, Lieven}, 
journal = {Cambridge University Press}, 
pages = {716}, 
language = {English}, 
month = {03}
}

@inproceedings{wang_industrial-strength_2003,
	address = {Baltimore, Maryland, USA},
	title = {An Industrial-Strength Audio Search Algorithm},
	language = {en},
	booktitle = {Proc. 4th {Int.} {Conf.} {Music} {Information} {Retrieval}},
	author = {Wang, Avery Li-Chun},
	month = oct,
	year = {2003},
}

@inproceedings{crocco2015room,
  title={Room impulse response estimation by iterative weighted l 1-norm},
  author={Crocco, Marco and Del Bue, Alessio},
  booktitle={2015 23rd European Signal Processing Conference (EUSIPCO)},
  pages={1895--1899},
  year={2015},
  organization={IEEE}
}

@misc{shazam2025,
  title        = {Shazam---{M}usic Discovery, Charts \& Song Lyrics},
  howpublished = {\url{https://www.shazam.com/}},
  note         = {Accessed: 15 Sep. 2025},
  year         = {2025},
}

@article{diamond2016cvxpy,
  author  = {Steven Diamond and Stephen Boyd},
  title   = {{CVXPY}: {A} {P}ython-embedded modeling language for convex optimization},
  journal = {J. Machine Learning Research},
  year    = {2016},
  volume  = {17},
  number  = {83},
  pages   = {1--5},
}

@misc{suno2025,
  title        = {{Suno AI}},
  howpublished = {\url{https://suno.com/}},
  note         = {Accessed: 15 Sep. 2025},
  year         = {2025}
}

\end{document}